\input harvmac
\input epsf

\def\nup#1({Nucl.\ Phys.\ $\bf {B#1}$\ (}
\noblackbox
\Title{\vbox{
\hbox{HUTP-99/A018}
\hbox{\tt hep-th/9903226}
}}{Conformal Approach to Particle Phenomenology}
\bigskip
\centerline{Paul H. Frampton $^{(1)}$ and Cumrun Vafa $^{(2)}$}
\bigskip
\centerline{$^{(1)}$ Department of Physics and Astronomy}
\centerline{University of North Carolina, Chapel Hill, NC 27599-3255}
\smallskip
\centerline{$^{(2)}$ Jefferson Laboratory of Physics}
\centerline{Harvard University, Cambridge, MA 02138}

\vskip .3in

We propose the existence of a
non-supersymmetric conformal field theory softly broken at
the TeV scale as a new mechanism
for solving the hierarchy problem.
We find the imposition of conformal invariance
to be very restrictive with many predictive
consequences, including severe restrictions on the field content,
 the number of families as well as on the structure of
inter-family
Yukawa couplings. A large
class of potentially conformal non-supersymmetric theories 
are considered and some general predictions are made about the 
existence of a rich spectrum of color and weak multiplets
in the TeV range.

\Date{March 1999}
\newsec{Introduction}
One of the basic facts about particle physics is the existence
of (at least) two very different scales: The Planck scale 
$\sim 10^{19} GeV$
and the weak scale $\sim 10^2 GeV$. The existence of these
two very different scales and their stability under quantum
correction is generally called the hierarchy problem. In fact there
are two separate issues here:  One is the existence
of two very different scales, and the other one is the stability
of such different scales under quantum corrections.  
As an example of the latter point generically
the mass of the scalars should be artificially fine tuned
to avoid getting the Planck scale masses by quantum corrections.
  One approach
to solving both aspects of the hierarchy problem is to postulate the existence
of supersymmetry which remains unbroken all the way down to
near the weak scale. Supersymmetry is responsible
for cancellation of many quantum corrections and this at least
makes the hierarchy technically possible.  The generation
of widely different scales can also be explained
if one assumes non-perturbative effects to be responsible.
Indirect evidence
for supersymmetry is mounting. These include the unification
of the gauge coupling constants of electroweak and strong 
forces in the minimal supersymmetric extension of the standard
model (but their lack of unification with the standard model
particle content) and
the successful prediction of the top quark mass. More theoretic
arguments for supersymmetry have also been advanced, for example
by appealing to its natural appearance in string theory.

Despite all this, there are some 
puzzles even in the
supersymmetric context including the $\mu$-problem
(i.e. why the Higgs mass is at the TeV scale as opposed
to the Planck scale--in other words the first aspect
of hierarchy problem)
and the issue of proton stability. 
More fundamentally breaking supersymmetry
would seem to lead to an unacceptably large value
for the cosmological constant.
Also the theoretic motivation
for supersymmetry as being fundamental for string theory
is questionable.  In fact one does have solutions
of string theories without any supersymmetry (see
e.g. \ref\kks{S. Kachru, J. Kumar and E. Silverstein,
``Vacuum Energy Cancellation in a Non-supersymmetric
String,'' hep-th/9810129.}\ for non-supersymmetric
solutions with zero (or small) cosmological constant).  Even in the
supersymmetric
compactifications one will have to deal with strings
in non-supersymmetric contexts (for example when one
considers thermal backgrounds).  For these reasons, as well
as because the issue is such a fundamental one for physics,
one should attempt alternative routes for solving the
hierarchy problem.

This paper is devoted to studying the possibility
of replacing supersymmetry
with conformal invariance near the
TeV scale in an attempt to solve the hierarchy problem.
We will show that this idea is in principle possible and we
present some toy models.
We will mainly address the issue of the stability
of the existence of two scales, and do not specifically
discuss the mechanisms of breaking conformal invariance
which should give rise to two very different scales.

We find that the principle of conformal
invariance is more rigid than supersymmetry in that in
many examples it predicts the number of generations as well
as a rich structure for Yukawa couplings among the various families.
This inter-family rigidity is a welcome feature
of the conformal approach to particle phenomenology.  In fact
it is challenging to come up with a conformal model
which satisfies all the known properties of the standard model.

The organization of this paper is as follows:  In the next
section we motivate the basic idea of considering conformal
theories for particle phenomenology.   In section 3 
we review a large class of non-supersymmetric
quantum field theories which are believed to lead
to conformal theories.  In fact we present further arguments,
using string dualities, supporting the conjecture
that the theories in question are conformal.
In section 4 we consider some aspects of model building using
the quantum field theories discussed in section 3.  In section 5
we end with some concluding remarks.

\newsec{Conformal Invariance as a Solution to the Hierarchy Puzzle}

The weak scale and the QCD scale, as well as the masses of
observed quarks and leptons are all so small compared to the Planck
scale, that it is reasonable to believe that in some approximation
they are exactly zero.  If so, then the quantum field theory which
would be describing the massless fields should be a conformal theory
(as it has no mass scales).  This simple observation suggests
that the gauge particles and the quarks and leptons, together
with some yet unseen degrees of freedom may combine to give
a quantum field theory with non-trivial
realization of conformal invariance.  In such a scenario
the fact that there are no large mass corrections follows by the condition
of conformal invariance.  In other words `t Hooft's naturalness condition is
satisfied, namely in the absence of masses there is an enhanced
symmetry which is conformal invariance.  We thus imagine the actual theory
to be given by an action
$$S=S_0+\int d^4 x \ \alpha_i O_i$$
where $S_0$ is the lagrangian for the conformal field theory in question,
and $O_i$ are certain operators of dimension less than 4, breaking
conformal invariance softly. The $\alpha_i$ represent
the ``mass'' parameters.  Their mass dimension is $4-\Delta_i$ where
$\Delta_i$ is the dimension of the field $O_i$ at the conformal point.
Note that the breaking should be soft, in order for the idea
of conformal invariance to be relevant for solving the hierarchy
problem.  This requires that the operators $O_i$ have dimension less than
4.

Let $M$ denote the mass scale determined by the parameters
$\alpha_i$.  This is the scale at which the conformal invariance
is broken. In other words, for energies $E>>M$ the couplings will not run
while they start running for $E<M$.  We will assume that $M$ is sufficiently
near the TeV scale in order to solve the hierarchy problem using
conformal invariance.  

This is our basic setup.  The main ingredient to fill is
to supply examples for conformal theories and see how close
can we come to the standard model.  In the next section
we review a large class of non-supersymmetric theories
in 4 dimensions which have been argued to be conformal
at least to the leading order in the large $N$
expansion.  We extend these arguments (making some
plausible assumptions) to finite
order in $N$. In section 4 we apply them to model building.

\newsec{Examples of Conformal Theories in 4 Dimensions}

In this section we review the construction
of a large class of quantum field
theories
in 4 dimensions, one for each discrete subgroup of $SU(4)$ and each
choice of integer $N$, motivated from string theory considerations
\ref\ks{S. Kachru and E. Silverstein, Phys. Rev. Lett. {\bf 80} (1998)
4855, hep-th/9802183.}\ref\lnv{A. Lawrence, N. Nekrasov and C. Vafa,
Nucl. Phys. {\bf B533} (1998) 199, hep-th/9803015.}.
It has been
proven that these theories
 have vanishing beta function to leading order
in $N$
using `t Hooft's planar diagrams embedded in string theory
\ref\bkv{M. Bershadsky, Z. Kakushadze and C. Vafa,
Nucl. Phys. {\bf B523} (1998) 59, hep-th/9803076.}\ or its translation to Feynman diagrams
\ref\bj{M. Bershadsky and A. Johansen,
Nucl. Phys. {\bf B536} (1998) 141, hep-th/9803249.}.  Below we argue for the existence of
at least one fixed point even for finite $N$ (under some technical
assumptions).  The vanishing of the beta function at large $N$
was also argued in \ks\ using AdS/CFT correspondence.

We will now describe the prescription for constructing
the above mentioned gauge theories.  We refer the reader
to \ks\lnv\ for the motivation for the prescription.  Roughly
speaking what the prescription does is to start with an $N=4$
gauge theory and get rid of some fields in the theory and identify
some of the other ones together in such a way that the resulting
theory is conformal. One ends up with theories with equal
number of bosons and fermions (if one includes the $U(1)$ parts)
but with bosons and fermions in different representations of the gauge group.
In particular one typically ends up with a non-supersymmetric theory.
The prescription is as follows:  We start with 
 $\Gamma\subset SU(4)$ which denotes a discrete subgroup of $SU(4)$
 (the global symmetry for $N=4$ supersymmetric gauge theory).
Consider irreducible representations of $\Gamma$.  Suppose there
are $k$ irreducible representations $R_i$, with dimensions $d_i$ with
$i=1,...,k$.  The gauge theory we construct has gauge symmetry
$$SU(N d_1)\times SU(Nd_2)\times ...\times SU(Nd_k)$$
for an arbitrary choice of integer $N$.
The fermions in the theory are given as follows.  Consider the 4 dimensional
representation of $\Gamma$ induced from its embedding in $SU(4)$.  It
may or may not be an irreducible representation of $\Gamma$. We consider
the tensor product of ${\bf 4}$ with the representations $R_i$:
\eqn\fdt{{\bf 4}\otimes R_i=\oplus_j n_i^jR_j}
The chiral fermions are in bifundamental representations
$$(1,1,..,{\bf Nd_i},1,...,{\overline {{\bf Nd_j}}},1,..)$$
with multiplicity $n_i^j$ defined above.  For $i=j$ the
above is understood in the sense that we obtain
$n_i^i$  adjoint fields plus $n_i^i$ neutral
fields of $SU(Nd_i)$.  
  Note that we can equivalently view
$n_i^j$ as the number of trivial representations in the tensor product
\eqn\coun{({\bf 4}\otimes R_i\otimes R_j^*)_{trivial}=n_i^j}
The asymmetry between $i$ and $j$ is manifest in the above
formula. Thus in general we have
$$n_i^j\not= n_j^i$$ and so the theory in question
is in general a chiral theory.  However
if $\Gamma$ is a real
subgroup of $SU(4)$, i.e. if ${\bf 4}={\bf 4}^*$ as far as
$\Gamma$ representations are concerned, 
then we have by taking the complex
conjugate of \coun :
$$n_i^j=({\bf 4}\otimes R_i\otimes R_j^*)_{trivial}=
({\bf 4}\otimes R_i\otimes R_j^*)^*_{trivial}=$$
\eqn\impo{({\bf 4}^*\otimes R_i^*\otimes R_j)_{trivial}=
({\bf 4}\otimes R_i^*\otimes R_j)_{trivial}=n_j^i.}
So the theory is chiral if and only if ${\bf 4}$ is a complex
representation of $\Gamma$, i.e. if and only if ${\bf 4}\not={\bf 4}^*$
as a representation of $\Gamma$.
If $\Gamma$ were a real subgroup of $SU(4)$ then
$n_i^j=n_j^i$. 

If $\Gamma$ is a complex subgroup, the theory is chiral, but
it is free of gauge anomalies.  To see this note that
the number of chiral fermions in the
fundamnetal representation of each group $SU(Nd_i)$ plus $Nd_i$
times the number of chiral fermions in the adjoint representation is given
by
\eqn\nufu{\sum_j n_i^j Nd_j=4 Nd_i}
(where the number of adjoints is given by $n_i^i$).
Similarly the number of anti-fundamentals plus $Nd_i$ times
the number of adjoints is given by
\eqn\anti{\sum_j n_j^i Nd_j=\sum Nd_j(4\otimes R_j\otimes R_i^*)_{trivial}=
\sum Nd_j(4^*\otimes R_j^*\otimes R_i)_{trivial}=4 Nd_i}
Thus, comparing with \nufu\ we see that
 the difference of the number of chiral fermions
in the fundamental and the anti-fundamental representation
is zero (note that the adjoint representation is real and does
not contribute to anomaly). Thus each gauge group is anomaly free.

In addition to fermions, we have bosons, also in the bi-fundamental
represenations.  The number of bosons $M_i^j$ in the bi-fundamental representation
of $SU(Nd_i)\otimes SU(Nd_j)$ is given by the number of $R_j$ representations
in the tensor product of the representation ${\bf 6}$ of $SU(4)$
restricted to $\Gamma$ with the $R_i$ representation.  Note that
since ${\bf 6}$ is a real representation we have
$$M_i^j=(6\otimes R_i\otimes R_j^*)_{trivial}=(6\otimes R_i^*\otimes
R_j)_{trivial}=M_j^i$$
In other words for each $M_i^j$ we have a {\it complex} scalar
field in the corresponding bi-fundamental representation, where complex
conjugation will take us from the fields labeled by $M_i^j$ to $M_j^i$.

The fields in the theory are natually summarized by a graph, called
the quiver diagram \ref\moord{M.R. Douglas and G. Moore,
``Quivers and ALE instantons,'' hep-th/9603167.}, 
where for each gauge group $SU(Nd_i)$ there
corresponds a node in the graph, for each chiral fermion in the
representation $(Nd_i, {\overline {Nd_j}})$, $n_i^j$ in total,
 corresponds a directed
arrow from the $i$-th node to the $j$-th node, and for each complex
scalar in the bifundamental of $SU(Nd_i)\times SU(Nd_j)$, $M_i^j$ in
total, corresponds
an {\it undirected} line between the $i$-th node and the $j$-th node
(see Fig. 1).

\subsec{Interactions}
The interactions of the gauge fields with the matter is fixed by
the gauge coupling constants for each gauge group. The inverse coupling
constant squared for the $i$-th group combined with the theta angle
for the $i$-th gauge group is 
$$\tau_i=\theta_i +{i\over 4\pi g_i^2}={d_i \tau \over |\Gamma|}$$
where $\tau=\theta+{i\over 4 \pi g^2}$ 
is an arbitrary complex parameter independent
of the gauge group and $|\Gamma|$ denotes the number of elements
in $\Gamma$. 

There are two other kinds of interactions: Yukawa
interactions and quartic scalar field interactions.
The Yukawa interactions are in 1-1 correspondence with
triangles in the quiver diagram with two directed
fermionic edges and one undirected scalar edge, with compatible directions
of the fermionic edges (see Fig. 1):
$$S_{Yukawa}={1\over 4\pi g^2}\sum_{directed \ triangles}
d^{{abc}}{\rm Tr}\psi_{ij^*}^a\phi_{jk^*}^b \psi_{ki^*}^c$$
where $a,b,c$ denote a degeneracy label of the corresponding
fields. $d^{abc}$ are flavor dependent numbers determined by
Clebsch-Gordan coefficients as follows:  $a,b,c$ determine elements $u,v,w$
(the corresponding trivial representation)
in ${\bf 4}\otimes R_i\otimes R_j^*$, ${\bf 6}\otimes R_j\otimes R_k^*$
and ${\bf 4}\otimes R_k\otimes R_i^*$. Then 
$$d^{abc}=u\cdot v\cdot w$$
where the product on the right-hand side corresponds
to contracting the corresponding representation indices for $R_m$'s
with $R_m^*$'s as well as contracting the
 $({\bf 4}\otimes {\bf 6}\otimes {\bf 4})$ according to the unique
 $SU(4)$ trivial representation in this tensor product.

 Similarly the quartic scalar interactions
are in 1-1 correspondence with the 4-sided polygons
in the quiver diagram, with each edge corresponding to an undirected
line (see Fig. 1).  We have
$$S_{Quartic}={1\over 4\pi g^2}\sum_{4-gons}f^{abcd}\Phi_{ij^*}^a
\Phi_{jk^*}^b\Phi_{kl^*}^c\Phi_{li^*}^d$$
where again the fields correspond to lines $a,b,c,d$ which in turn
determine an element in the tensor products of the form
${\bf 6}\otimes R_m\otimes R_n^*$.  $f^{abcd}$ is obtained
by contraction of the correponding element
as in the case for Yukawa coupling and also using a $[\mu ,\nu][\mu ,\nu]$
contraction in the ${\bf 6}\otimes {\bf 6}\otimes {\bf 6}\otimes {\bf 6}$ part
of the product.

\bigskip
\epsfxsize 2.truein
\epsfysize 2.truein
\centerline{\epsfxsize 2.truein \epsfysize 2.truein\epsfbox{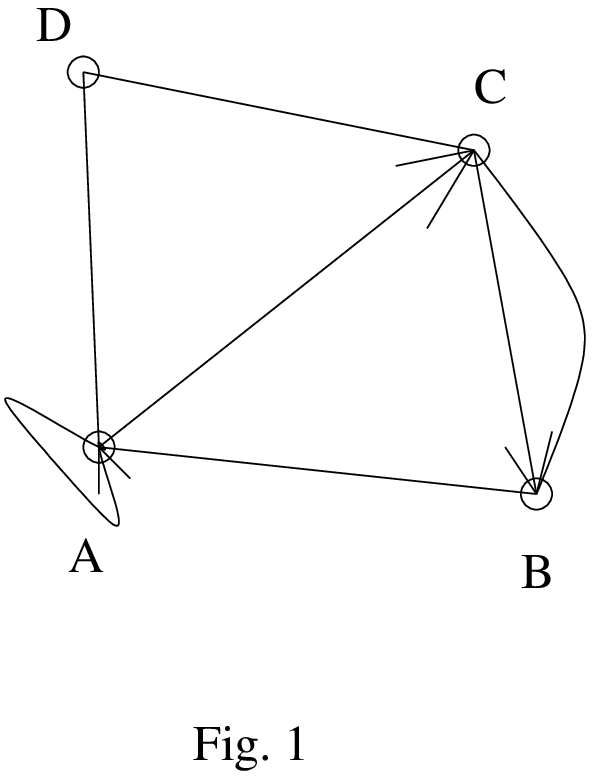}}
\leftskip 2pc
\rightskip 2pc
\noindent{\ninepoint\sl \baselineskip=8pt {\bf Fig.1}: {\rm A part
of the Quiver Diagram. The edges with arrow represent fermionic
matter and the edges without arrows represent scalar matter.
The Yukawa coupling in this diagram comes from the ACB
triangle and the quartic coupling comes from the ABCDA square.
The edges whose beginning and end are the same point correspond
to adjoint matter.}}

\subsec{Conformality}
As mentioned before it has been shown that the above
theories are conformal to leading order in a
$1/N$ expansion for arbitrary values of $\tau$.  The question
is whether this can be pushed to go beyond the leading order in 
the $1/N$ expansion.
If $\Gamma $ resides in an $SU(2)$ subgroup of $SU(4)$
then one obtains an $N=2$ superconformal theory and 
the conformality is automatic for finite $N$ as well, by
non-renormalization theorems of $N=2$. In fact it has
been shown that all the finite $N=2$ theories with
matter representations in the bi-fundamental representation
of the groups are captured by the above construction
\ref\kmv{S. Katz, P. Mayr and C. Vafa,
Adv. Theor. Math. Phys. {\bf 1} (1998) 53, hep-th/9706110.}, and are thus in 1-1 correspondence
with discrete subgroups of $SU(2)$.  In particular
the corresponding reduced quiver diagram corresponds,
in this case, to the affine $A-D-E$ Dynkin diagrams 
(for cyclic, dihedral and exceptional subgroups
of $SU(2)$ respectively) with
each link of the Dynkin diagram corresponding to hypermultiplet
matter, and the rank of each gauge group being proportional
to the corresponding Dynkin numbers. In this case one
can actually change the coupling of each gauge group independently
and still obtain a conformal theory.
If $\Gamma$
resides in an $SU(3)$ subgroup, 
one obtains an $N=1$ supersymmetric theory.  In this case
one can use
$N=1$ non-renormalization theorems, along the lines
of \ref\stra{R.G. Leigh and M.J. Strassler,
Nucl. Phys. {\bf B447} (1995) 95, hep-th/9503121.}\ to prove for the existence of
conformal theories to all orders.
In these cases there would be at least a one dimensional
line of fixed points.

If $\Gamma$ resides in the full $SU(4)$ group, then
we have a non-supersymmetric theory, and it is natural
to ask to what extent one can expect conformal theories for finite
values of $N$.  It has been conjectured in \lnv\ that this
continues to be the case, at least for some choice of
the coupling constants.  Here we will give a plausibility
argument for this conjecture.  At the leading order
in $1/N$, as we have indicated above, all the coupling
constants in the Lagrangian are determined by a single
coupling constant $\tau$, and we thus effectively
have only one coupling constant.  Under the Renormalization
group flow it is not necessarily true that all the couplings
in the theory
will still be determined by a single coupling constant; however
let us assume this continues to be correct.
These theories have an additional property that has not been
mentioned so far, and that is the $\tau \rightarrow -1/\tau$ 
strong-weak duality.  This duality exchanges $4 \pi g^2\leftrightarrow
 1/4 \pi g^2$ (at $\theta =0$).  This follows from their
 embedding in the type IIB string theory which enjoys
 the same symmetry.  In fact this gauge theory {\it defines}
 a particular type IIB string theory background
 \ks\ and so this symmetry must be true for the gauge theory
 as well.  In the leading order in $N$ the beta function vanishes.
 Let us assume at the next order there is a negative beta function,
 i.e., that we have an asymptotically free theory.  Then
 the flow towards infrared increases the value of the coupling
 constant. Similarly, by the strong-weak duality, the flow towards
 the infrared at large values of the coupling constant must
 decrease the value of the coupling constant.  Therefore
 we conclude that the beta function must have at least one
 zero for a finite value of $g$.  
 
 This argument is not rigorous for three reasons:  One is that
 we ignored the flow for the $\theta $ angle.  This can be
 remedied by using the fact that the moduli space is the upper
 half-plane modulo $SL(2,Z)$ which gives rise to a sphere
 topology and using the fact that any vector field has a zero
 on the sphere (``you cannot comb your hair on a sphere'').
 The second reason is that we assumed asymptotic freedom
 at the first non-vanishing order in the large $N$ expansion.
 This can in principle be checked by perturbative techniques
 and at least it is not a far-fetched assumption.
 More serious, however, is the assumption that there is effectively
 one coupling constant. It would be interesting to see if one
 can relax or verify this assumption, which is valid at large $N$.
 
 \newsec{Applications to Particle Phenomenology}
 
 In section 2 we have outlined the basic strategy
 of connecting conformal theory to particle phenomenology
 and in section 3 we have given a large list of quantum field
 theories which are potentially conformal. In this section we
 would like to indicate by general arguments and also
 through examples, how particle
 phenomenology may arise from such a picture.  Even though
 there presumably are many more conformal theories without
 supersymmetry, we will mainly concentrate on the ones that we
 discussed in section 3.
 For an extension of these models, see \ref\kakus{
 Z. Kakushadze, Nucl. Phys. {\bf B544} (1999) 265,
hep-th/9808048\semi
 Phys. Rev. {\bf D59} (1999) 045007, hep-th/9806091.}.  Hopefully more
examples
 will be known
 in the near future, which one may apply to particle phenomenology.
 
 As discussed in section 2 we assume that the Lagrangian
 describing the particles and their interaction is nearly
 conformal, i.e. it is a soft-breaking of a conformal theory.
 In the examples provided in section 3 the soft breaking terms
 would involve arbitrary additions to the Lagrangain involving
 quadratic scalar mass terms, fermionic mass terms and cubic
 scalar mass terms consistent
 with gauge interactions.  From the perspective of quiver
 diagrams this means considering 2-gons with two compatibly directed
 edges (corresponding to fermion mass term) or 2-gons with two
 undirected edges, corresponding to scalar mass terms, or triangles
 with three undirected edges, corresponding to cubic scalar couplings:
\eqn\deact{ S=S_0+\int \alpha_{ab}{\rm Tr} \Psi_{ij^*}^a \Psi_{ji^*}^b
 +\alpha_{cd}^2{\rm Tr} \Phi_{ij^*}^c\Phi_{ji^*}^d+
 \alpha_{efg}{\rm Tr} \Phi_{ij^*}^e\Phi_{jk^*}^f\Phi_{ki^*}^g+c.c.}

 Depending on the sign of quadratic terms
 for the scalars in the above action, the conformal
 breaking terms in \deact\ could induce 
 gauge symmetry breaking using the Higgs mechanism.
 Consider for example two gauge groups $SU(Nd_i)\times SU(Nd_j)$
 and let us suppose that the above terms are such that $\langle 
 \phi_{ij^* }\rangle \not=0$. For the sake of example
 let us assume $d_i=d_j=d$. Then we can represent
 the expectation value of $\phi_{ij^*}$ as a square
 matrix with diagonal entries.  Depending on the eigenvalues 
 of the matrix we get various patterns of gauge symmetry
 breaking. For example if we have 2 equal non-vahnishing eigenvalues and
 the rest zero we get the breaking pattern
 $$SU(Nd)\times SU(Nd)\rightarrow SU(2)_{diagonal}\times U(1)\times
 SU(Nd-2)\times SU(Nd-2).$$
 If we have more scalars in the bifundamentals
 of the $SU(Nd)\times SU(Nd)$
 we can break the groups even further, by various
 alignments of the expectation values.   Thus in general we can have
 a rich pattern of gauge symmetry breakings.
 
\subsec{Some General Predictions}
We would like to discuss how $SU(3)\times SU(2)\times U(1)$
standard model can be imbedded in the conformal theories under discussion.
In other words we consider some embedding
$$SU(3)\times SU(2)\times U(1)\subset \otimes_i SU(Nd_i)$$
in the set of conformal theories discussed in section 3.  Each gauge
group of the standard model may lie in a single $SU(Nd_i)$ group
or in some diagonal subgroup of a number of $SU(Nd_i)$ gauge
groups in the conformal theory.
The first fact to note, independently of the embeddings, is that
the matter representations we will get in this way are severly restricted.
This is because in the conformal theories we only have bi-fundamental
fields (including adjoint fields), and thus any embedding of the standard model in the conformal
theories under discussion will result in matter in bi-fundamentals
(including adjoints), and no other representation. 
 For example we cannot have a matter field tranforming
 according to representation of
the form $({\bf 8},{\bf 2})$ of $SU(3)\times SU(2)$. That we can
have only fundamental fields or bi-fundamental fields is a strong restriction 
 on the matter content of the standard model which in
fact is satisfied and we take it as a check (or evidence!) for the
conformal approach to phenomenology.  The
rigidity of conformal theory in this regard can be compared
to other approaches, where typically we can have various kinds
of representations.

Another fact to note is that there are no $U(1)$ factors in the
conformal theories (having charged $U(1)$ fields is in conflict
with conformality) and in particular the existence of quantization
of hypercharge is automatic in our setup, as the $U(1)$ has to be
embedded in some product of $SU$ groups.  This is 
the conformal version of the analogous statement
in the standard scenarios to unification, such as $SU(5)$
GUT.  
 
 The rigidity of conformal invariance goes much further.
 Embedding the standard model in the conformal theories
 predicts a rich spectrum of additional unobserved
 particles charged under $SU(3)$ and $SU(2)$.  These extra
 particles should have mass in the TeV range (the conformal
 breaking scale) in order for conformal invariance
 to be relevant for the resolution of the hierarchy puzzle. 
   The lower
 bound on the number of extra particles arises if we assume
 each group of the standard model is embedded in a single $SU(Nd_i)$
 group of the same size.  This is because the number of extra fermions
 transforming under fundamental representation 
  (plus $Nd_i$ times the number of adjoint fields) is
  given by $4 Nd_i$ for each $SU(Nd_i)$ and taking a diagonal
 subgroup of them will only give rise to more matter fields. Moreover
 the minimal extra matter will come by assuming $Nd_i$ is equal
 to $3$ and $2$ for the $SU(3)$ and $SU(2)$ gauge groups respectively.
 To obtain the lower bound on the number of extra color
 charged matter fields we thus assume that one of the
 gauge groups in $S_0$ is the $SU(3)$ color.  Then from the
 constructions discussed in the previous section we see
 that the number of the triplets $N_3$ plus $3$ times
 the number of adjoints $N_8$ is bounded by
 $$N_3+3 N_8\geq 4\cdot 3=12$$
 Subtracting the observed number of quarks $N_3=6+\Delta N_3$
 we see that
 $$\Delta N_3+ 3 N_8\geq 6$$
 In other words, if all the extra color charged
 fermions were in the fundamental representations, we would be
 predicting at least 6 more quarks in the TeV range. Of course
 the actual number of families will depend on the rest
 of the gauge groups.  For example if the breaking
 goes through an $SU(3)\times SU(3)\times SU(3)$ gauge
 group, as in the trinification
 scenario \ref\glet{A.De Rujula, H. Georgi and S.L. Glashow,
 ``Trinification of all elementary particle forces,'' presented
 by S.L. Glashow at the Fifth Woskshop on Grand Unification,
 Brown University, 1984.}\
  each family would contain 3 quarks (one of which
 must be much heavier than the other two to be phenomenologically
 realistic) and in the case of the lower
 bound we would end up with 4 families.  
 Similarly we predict the existence of extra colored scalars.
 If we denote the number of complex triplet scalars by $M_3$
 and the number of adjoint scalars by $M_8$ we would predict
 $$M_3+3M_8\geq 6\cdot 3=18$$
 Going over the same exercise for the $SU(2)$ group
 we would predict the following bound on the extra
 weak charged fermions and bosons in the TeV range:
 $$\Delta N_2+4 N_3\geq 4$$
 $$\Delta M_2+2 M_3\geq 11$$
 (where we subtracted 12 from $N_2$ and $1$ from $M_2$ corresponding
 to the matter content in the standard model).
 These bounds clearly demonstrate the rigidity of the
 conformal approach to phenomenology.
 
 As far as other rigidities
 that the conformal assumption introduces, 
 we should note that the structure of the Yukawa
 and quartic couplings is untouched by the soft-breaking 
 terms and is determined
 completely in terms of the fixed point values at the conformal
 point given in $S_0$.  As we saw in the last section, these
 Yukawa terms have rich flavor dependent structure which is
 dictated by conformal invariance.  This is in sharp
 contrast to standard phenomenology or the MSSM where
 the Yukawa couplings are put in by hand.  We thus see that
 conformal symmetry is far more rigid than supersymmetry.
 
 \subsec{Coupling Unification}
 One of the successful ideas about GUTs is the
 prediction of the meeting of the three gauge coupling constants
 of the standard model at a high energy scale ($\sim 10^{16} GeV$)
  which is a beautiful feature of
 the minimal supersymmetric
 extension of the standard model.  In the case at
 hand, however we are assuming the existence of a conformal
 theory in the TeV scale. Beyond the conformal symmetry
 breaking scale the gauge group $\otimes SU(Nd_i)$ is restored
 and their 
 couplings will not run. Even though the gauge
 groups do not unify into a single gauge group, 
 recall from section 3 that the coupling
 constants of the gauge groups are related to each other,
 which is also a feature of GUT theories.  Matching the coupling
 constant of the $SU(3)\times SU(2)\times U(1)$ gauge theories at
 the conformal breaking scale with the fixed point values
 for the coupling constants at the conformal point,
 we see that the coupling constants of the $SU(3)$ and $SU(2)$
 and $U(1)$ will be strongly correlated, though not necessarily
 equal.  Even if the gauge coupling constants of the $SU(Nd_i)$ groups
 were all equal (which is in general not true) we cannot
 deduce that the gauge couplings of the $SU(3)\times SU(2)\times U(1)$
 gauge theories are equal at the conformal breaking scale.  In particular
 if $SU(3)$ embeds into a single $SU(Nd_i)$ gauge group, and $SU(2)$
 in two such groups and $U(1)$ in 6 such groups, we would have
 gotten the ratios of the coupling constants
 $\tau_3/\tau_2/\tau_1\sim 1/3/6$ which would be close to the observed
 ratios in the weak scale.  We thus see that in the present
 scenario the coupling constants are strongly correlated,
 yet they do not have to be equal unlike the case in the standard
 GUTs.

 \subsec{Some Examples}
 In this section we will consider some illustrative examples. 
 In order to specify an $S_0$ in the class of Lagrangian
 we considered in the previous section, we need to start
 with a discrete subgroup $\Gamma \subset SU(4)$.  Moreover
 if we wish to have a chiral theory we should consider
 complex subgroups of $SU(4)$.  The simplest (though not
 necessarily the most interesting) subgroups are the cyclic ones.
 Let us look for a non-supersymmetric chiral model.
 The $Z_2$ subgroup, consisting of $4(-1)$ eigenvalues, is real
 and so is not a chiral theory.  For the $Z_3$ subgroup 
 we have two choices: one is three equal eigenvalues and one eigenvalue
 1, which corresponds to an $N=1$ supersymmetric chiral theory
 (for the rank 3 case we would get 3 families of the supersymmetric
 version of trinification without any Higgs multiplets).  Another
 choice of $Z_3$ consists of two pairs of conjugate eigenvalues.
 This is a real representation and is not chiral. The next case
 is $Z_4$ and the only complex representation without supersymmetry
 is the embedding with four eigenvalues of $i$. This gives a
 gauge group $SU(N)^4$ for some choice of $N$ with fermion
 and bosons given by the quiver diagram below:
 
 \bigskip
\epsfxsize 2.truein
\epsfysize 2.truein
\centerline{\epsfxsize 2.truein \epsfysize 2.truein\epsfbox{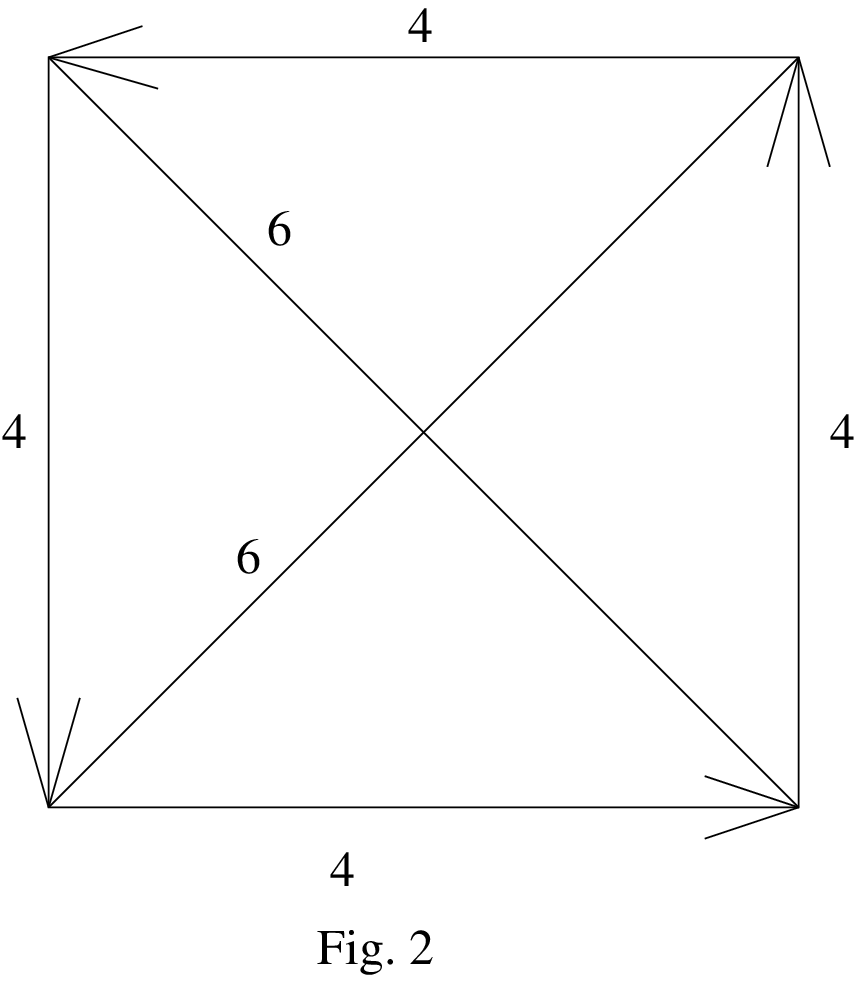}}
\leftskip 2pc
\rightskip 2pc
\noindent{\ninepoint\sl \baselineskip=8pt {\bf Fig.2}: {\rm 
The quiver diagram for the $Z_4$ theory.  Fermions
correspond to the edges and the diagonals correspond to
the scalars.  The labels next to the lines denote the degeneracy of
each line.}}

 The four gauge groups are identified with the vertices of the square.
 The fermions are four copies of the edges of the square and
 the scalars are 6 complex fields identified with the diagonals.
 The scalar content of this theory
 is too rigid to give a breaking pattern which could lead to the
  standard model spectrum.  The simplest non-supersymmetric
  chiral model which contains the spectrum of the standard
  model and can be broken to it is a $Z_5$ subgroup of $SU(4)$
  where we choose the four eigenvalues to be
  $ (\alpha,\alpha,\alpha,\alpha^2)$ where $\alpha ={\rm exp}(2\pi i/5)$.
  For the gauge group we take $SU(3)^5$ and it is convenient to identify
  them with 5 points on the circle, labeled in order by 0 through 4
  with the integers defined mod $5$.
  The theory at the conformal point has a $Z_5$ cyclic permutation
  of the gauge groups (along with other symmetries which we
  will not consider here).  There are three chiral fermions
  in the bifundamental of $(i,i+1)$ gauge groups and one
  bi-fundamental in the $(i,i+2)$ gauge groups where $i$ runs
  over all the gauge groups.  There are in addition 3 complex
  scalars in the bifundamental of the $(i,i+2)$ and 3 complex
  scalars in the bifundamental of $(i,i+3)$.  See  the quiver
  diagarm below.
  
  \bigskip
\epsfxsize 2.truein
\epsfysize 2.truein
\centerline{\epsfxsize 4.truein \epsfysize 4.truein\epsfbox{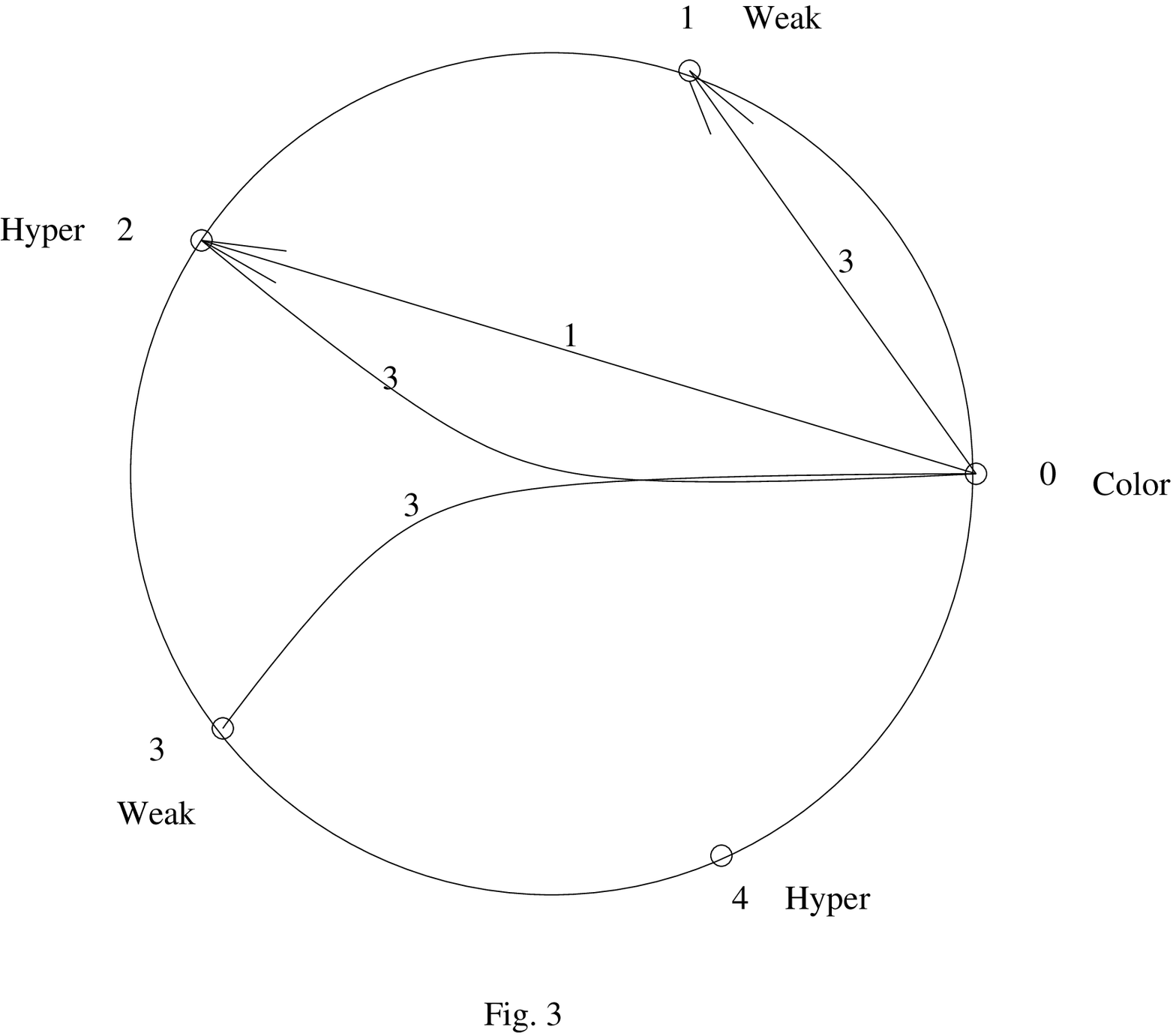}}
\leftskip 2pc
\rightskip 2pc
\noindent{\ninepoint\sl \baselineskip=8pt {\bf Fig.3}: {\rm 
The quiver diagram for the $Z_5$ conformal theory.
In order to simplify the diagram we have drawn
only the edges emenating from the 0 node.  The rest
are found by the cyclic symmetry of the theory under
the cyclic permutation (01234).  The $SU(3)$ color
is embeded in the gauge group at node 0, and
the $SU(2)$ weak is embedded in the diagonal subgroup
of nodes 1 and 3, and the third gauge group
in the trinification (which we have rather loosely called
hyper) comes from the diagonal embedding in the gauge
groups 2 and 4.}}

  We identify the $SU(3)$ color group with the gauge group at node
  $0$.  We assume that the conformal breaking deformation
  breaks the groups $(1,3)$ to a diagonal $SU(3)$ 
  (which contains the $SU(2)$ weak) and the groups
  $(2,4)$ to a diagonal $SU(3)$.  We end up with 3 families
  of fermions in the standard trinification representation
  $$({\bf 3},{\overline{\bf 3}},1)+(1,{\bf 3},{\overline{\bf 3}})+
  ({\overline{\bf 3}},1,{\bf 3})$$
  and one family in the conjugate representation. In addition
  we have one fermionic fields in the adjoint of the second and
  third $SU(3)$'s as well as 3 copies of $[(1,{\bf 3},{\overline
{\bf 3}})+c.c.]$
.  In addition there are a number
of scalars
  in various bi-fundamental represenations.  There are enough
  scalars to induce the breaking of the trinification
  group to the standard $SU(3)\times SU(2)\times U(1))$,
  which require only two bi-fundamental field in the last
  two $SU(3)$'s which is available.
  The point of this exercise was to show how the standard
  model may arise, and not to overemphasize this particular
  model.
 
 \newsec{Concluding Remarks}
 It should be clear from the various examples presented
 how restrictive the assumption of conformal invariance
 is for particle phenomenology. It is precisely for this reason
 that it is potentially useful as an organizing principle
 for particle physics.  
 
 Here we have concentrated on a large class of conformal
 field theories (which have been conjectured to be
 the complete list \lnv\ if one restricts attention to theories with
 bi-fundamental matter).  However there may well be other
 interesting conformal theories for particle phenomenology
 and this issue should be further investigated.
 Even within the class of models considered, we would need
 a complete classification of subgroups of $SU(4)$ and
 their representation ring--this is mathematically within
 reach, but has not been done yet (for the $SU(3)$ case 
 this is already known--see \ref\math{W.M. Fairbanks,
 T. Fulton and W.H. Klink, J. Math. Phys. {\bf 5} (1964) 1038.}\ref\mit{A.
 Hanany and Y.-H. He,``Non-abelian Finite Gauge Theories,'' hep-th/9811183.}
).  Some progress in this direction has been made
in \ref\plet{W. Plesken and M. Pohst, Math. Comp. {\bf 31} (1977) 552.}.
 
 A key issue to understand in connection with 
 applying the ideas of the present paper to particle
 physics is to identify natural mechanisms to softly
 break conformal invariance.  In particular one would
like to get an explanation of the generation
of a small scale compared to the Planck scale.
It is tempting to speculate that
 quantum gravity effects can play an interesting
 role here.
 
 Within the class of models
 introduced, we have found a lower bound
 on the number of new charged fields in the TeV
 scale.
 Thus if conformal symmetry is realized
 near the TeV scale,  we will be witnessing discovery of many
 new particles in the accelerators in the near future.
 
We would also like to thank
T. Banks,
H. Georgi, S. Glashow, Z. Kakushadze and E. Witten for valuable
discussions.
 
 The research of P.F. was supported in part by DOE grant
 DE-FG02-97ER41036.
The research of C.V. was supported in part by NSF grant PHY-92-18167.

 \listrefs

\end